\documentclass[11pt,a4paper]{article}
\pdfoutput=1

\usepackage{jheppub}

\usepackage{amsmath}
\usepackage{amssymb}
\usepackage{graphicx}

\usepackage{subfigure}

\newcommand{\ie}{{\it i.e.}}

\newcommand{\be}{\begin{equation}}
\newcommand{\ee}{\end{equation}}
\newcommand{\br}{\begin{eqnarray}}
\newcommand{\bea}{\begin{eqnarray}}
\newcommand{\eea}{\end{eqnarray}}
\newcommand{\er}{\end{eqnarray}}
\newcommand{\ba}{\begin{array}}
\newcommand{\ea}{\end{array}}
\newcommand{\bi}{\begin{itemize}}
\newcommand{\ei}{\end{itemize}}
\newcommand{\bn}{\begin{enumerate}}
\newcommand{\en}{\end{enumerate}}
\newcommand{\bc}{\begin{center}}
\newcommand{\ec}{\end{center}}

\newcommand{\hc}[2][]{#2^{\dagger #1}} 
\newcommand{\abs}[1]{|#1|} 

\def\gappeq{\mathrel{\rlap {\raise.5ex\hbox{$>$}}
{\lower.5ex\hbox{$\sim$}}}}

\def\lappeq{\mathrel{\rlap{\raise.5ex\hbox{$<$}}
{\lower.5ex\hbox{$\sim$}}}}

\title{Implications of the 125 GeV Higgs boson for scalar dark matter and for the CMSSM phenomenology}

\author[a]{Mario Kadastik,}
\author[a,b]{Kristjan Kannike,}
\author[a]{Antonio Racioppi}
\author[a]{and Martti Raidal}

\affiliation[a]{National Institute of Chemical Physics and Biophysics, Ravala 10,
Tallinn 10143, Estonia}
\affiliation[b]{Scuola Normale Superiore and INFN, Piazza dei Cavalieri 7, 56126 Pisa, Italia}

\emailAdd{mario.kadastik@cern.ch}
\emailAdd{kannike@cern.ch}
\emailAdd{antonio.racioppi@kbfi.ee}
\emailAdd{martti.raidal@cern.ch}

\abstract{We study phenomenological implications of the ATLAS and CMS hint of a $125\pm 1$~GeV Higgs boson for the  singlet, and
singlet plus doublet non-supersymmetric dark matter models, and for the phenomenology of the CMSSM. We show that in 
scalar dark matter models the vacuum stability bound on Higgs boson mass is lower than in the standard model 
and the 125~GeV Higgs boson is consistent with the models being valid up the GUT or Planck scale.
We perform a detailed study of the full CMSSM parameter space keeping the Higgs boson mass fixed to $125\pm 1$~GeV, and study in detail
the freeze-out processes that imply the observed amount of dark matter. After imposing all phenomenological constraints except for the muon $(g-2)_\mu,$ 
we show that the CMSSM parameter space is divided into well separated regions with distinctive but in general heavy sparticle mass spectra.
Imposing the $(g-2)_\mu$ constraint introduces severe tension between the high SUSY scale and the experimental measurements -- 
only the slepton co-annihilation region survives with potentially testable sparticle masses at the LHC. 
In the latter case the spin-independent DM-nucleon scattering cross section is predicted to be below detectable limit at the XENON100
but might be of measurable magnitude in the general case of light dark matter with large bino-higgsino mixing and unobservably large
scalar masses.}

\arxivnumber{1112.3647}

\begin{document}

\maketitle

\section{Introduction}
In the standard model (SM) of particle interactions the only unknown quantity is the Higgs boson 
mass~\cite{Higgs:1964ia,Guralnik:1964eu,Englert:1964et,Higgs:1964pj}.
Any assumption that fixes the Higgs boson quartic self-coupling at any scale $\Lambda$ implies a prediction
for the Higgs boson mass. Many models of that sort have been proposed in the past based on different
arguments of new physics beyond the SM. In general, the properties of the SM Higgs potential are among
 the best studied quantities in particle physics~(\cite{Xing:2011aa}; for a review and references see \cite{EliasMiro:2011aa}).

 Based on  data collected in 2011,  both the ATLAS and CMS experiments have published their results for
 searches for the SM-like Higgs boson~\cite{Chatrchyan:2012tx,ATLAS:2012si} confirming and improving their earlier claims~\cite{ATLAS-CONF-2011-163,CMS-PAS-HIG-11-032}
 for the inconclusive evidence of a signal of a $M_H=124$~GeV (CMS) or $M_H=126$~GeV (ATLAS) Higgs boson; we will assume that the mass is in this $M_H=125 \pm 1$~GeV range.
 The corresponding local significances of the excess in ATLAS and CMS are $3.5\sigma$ and $3.1\sigma$, respectively, 
 while the  global significances after taking into account the look-elsewhere-effect  are 
 $2.2\sigma$ and $2.1\sigma$.
 Although definitive confirmation of the observed evidence requires more data,  the LHC result motivates studies of fundamental
 scalars in particle physics and in cosmology.

If  the present  inconclusive evidence for $M_H \approx 125$~GeV Higgs boson will be confirmed, this result will have a profound
impact on building models beyond the SM and on their phenomenology. 
In the context of the SM, the Higgs boson mass 125~GeV is below the vacuum stability bound  $M_H>128$~GeV 
coming from the requirement of the SM validity up to the scale of gauge coupling unification $\Lambda_{\rm GUT}.$ 
Vanishing SM Higgs boson self-coupling $\lambda (\Lambda)=0$ below the GUT scale, $\Lambda<\Lambda_{\rm GUT},$ 
implies that  the fundamental scale of new physics related to electroweak symmetry breaking and, perhaps, to flavour generation, might be 
lower than the GUT scale. On the other hand, the Higgs boson mass $M_H \approx 125$~GeV may imply that there is new 
 physics beyond the SM not too far  from the electroweak scale that modifies the Higgs boson mass prediction.
The most popular such a framework is low energy supersymmetry (SUSY) that prefers a light Higgs boson. For SUSY
scenarios the lightest Higgs boson mass $M_H \approx 125$~GeV is unusually high, close to the upper bound in popular models, 
and implies a higher SUSY breaking scale than one expects from naturalness arguments. Clearly those arguments mean that the 
present hint for the Higgs boson mass requires re-assessment of  several  ``standard" concepts both in SUSY and in non-SUSY models.

The aim of this work is twofold. First, assuming that the Higgs boson mass is in the range $M_H=125\pm 1$~GeV,
we study the implications of this assumption on the vacuum stability in scalar dark matter (DM)  models.
In those models the DM and Higgs sectors are related via the Higgs portal and the scalar potentials are in general rather complicated. 
Due to many new self-interactions in the scalar sector, the SM Higgs quartic coupling renormalization is modified and  one might expect that 
the triviality $\lambda (\Lambda)=0$ may be achieved for higher values of $\Lambda.$ We show that this is indeed the case and the SM vacuum
stability results will be changed in the non-SUSY scalar DM models compared to the SM prediction. 
As a new result we show that in those scenarios the 125~GeV Higgs boson is consistent with the
vacuum stability  up to $ \Lambda_{\rm GUT}$ and, therefore, the scalar DM models do not require new fundamental
scales between TeV and  the GUT scales.

Second, a technically much more involved question is what is the implication of  the $M_H=125\pm 1$~GeV LHC result for SUSY predictions of
generating DM relic abundance, DM direct detection and for the LHC phenomenology. Generically such a heavy
Higgs boson requires rather heavy stops, \ie, a large SUSY breaking scale\footnote{In the context of the 125~GeV Higgs boson this point has
already been noted in \cite{Carena:2011aa,Moroi:2011aa,Moroi:2011ab,Draper:2011aa,Arbey:2011ab,Heinemeyer:2011aa,Li:2011ab,Baer:2011ab,Hall:2011aa,Arbey:2011aa}.}. 
This, in general,  implies a large fine tuning to obtain
the correct electroweak scale,  very fine tuned DM annihilation channels and poor prospects for discovering SUSY at the LHC.
We analyze those issues in detail in  the constrained minimal
supersymmetric standard model (CMSSM) and show that the requirements of $M_H=125\pm 1$~GeV and correct DM relic abundance
together select out parameter regions with  well defined sparticle spectra. We work out CMSSM predictions for DM direct detection cross sections
in those parameter regions. The most important new result of this paper is to predict sharp linear relationship between the
gluino, lightest stop and slepton masses  in the stop and slepton co-annihilation regions that are the only ones accessible to the LHC experiments.

 If, in addition, also the muon anomalous magnetic moment $(g-2)_\mu$ constraint is imposed on the CMSSM, only a tiny parameter region
is singled out that induces DM via the slepton co-annihilation channel. In this parameter space the LHC has a good chance to
observe gluinos and the lightest stop but the DM direct detection experiments like  XENON100 are predicted to obtain null result. 
In the other DM freeze-out channels that also predict the correct amount of DM  the situation might be an opposite -- only
TeV scale DM is observable in DM direct detection experiments while the heavy gluinos and scalars decouple from the spectrum.
We classify all  those possibilities and discuss their phenomenology.

In section~\ref{sec:scalars} we present results for models of the SM extended with scalars: a complex $SU(2)$ singlet, an inert doublet or both. In section~\ref{sec:cmssm} we give scans for CMSSM with both with and without the $(g-2)_\mu$ constraint. We conclude in section~\ref{sec:concl}.

\section{Scalar dark matter and vanishing Higgs self-coupling}
\label{sec:scalars}

\begin{figure}[t]
\centering
\includegraphics{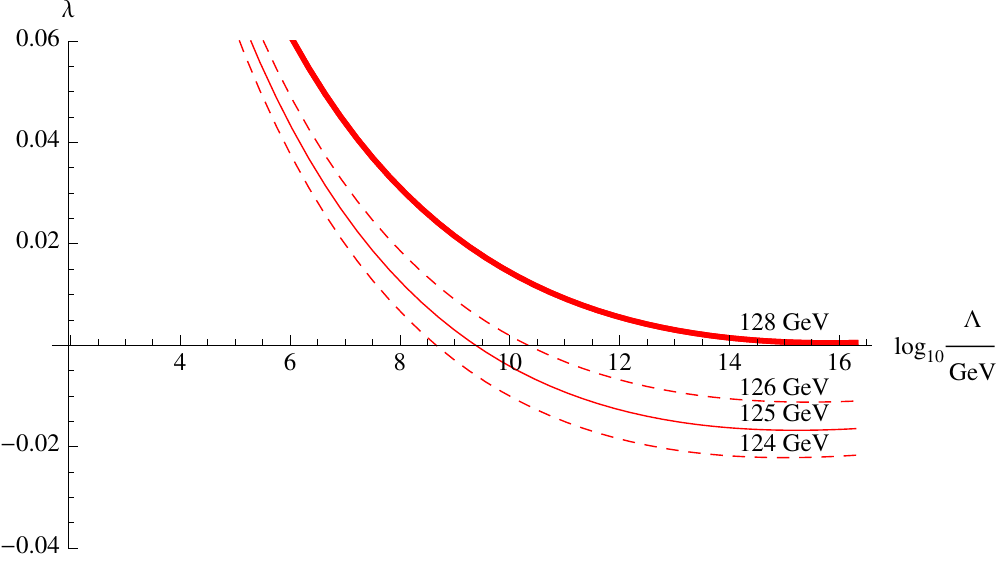}
\caption{Running of the SM Higgs boson self-coupling $\lambda$ for different Higgs boson masses at two loop level.}
\label{fig1}
\end{figure}

Triviality of the SM Higgs boson self-coupling, $\lambda=0,$ at some scale $\Lambda$ is an interesting possibility. 
From theoretical point of view this may indicate a scale where some new fundamental theory beyond the SM generates 
electroweak symmetry breaking and Higgs boson Yukawa couplings, \ie, flavour physics. 
From the phenomenological point of view this scale uniquely predicts the Higgs boson mass due to the evolution of the Higgs self-coupling via
renormalization group equations. Examples of this running at two loop level in the SM are presented in Fig.~\ref{fig1} for 
different values of the SM  Higgs boson masses as indicated in the figure. Our results agree with the recent works \cite{Xing:2011aa,EliasMiro:2011aa}.
This result shows that the LHC indications for the Higgs boson imply the triviality scale to be about $10^{10}$~GeV rather than 
the GUT scale $2.1\times 10^{16}$~GeV. Such a low scale can be associated with the seesaw scale \cite{Gell-Mann:1979kx,Yanagida:1979uq,Mohapatra:1979ia,Glashow:1979nm,Minkowski:1977sc} where neutrino masses are generated
rather than with the GUT scale.

The natural question to ask is that what happens to the vacuum stability in models with extended scalar sector?
Particularly interesting among those models are the scalar DM models that have been already addressed in the 125~GeV Higgs boson 
scenario~\cite{Djouadi:2011aa}.\footnote{Singlet fermion DM has also been studied \cite{Baek:2011aa}.}

\subsection{Scalar singlet model}

The simplest DM model is obtained by extending  the SM scalar potential with a real~\cite{McDonald:1993ex,Burgess:2000yq,Barger:2007im,Gonderinger:2009jp} or complex~\cite{Barger:2008jx} singlet scalar field. 
 In view of embedding this scenario into a GUT framework~\cite{Kadastik:2009dj}, 
 we study the complex singlet scalar $S = (S_H + i S_A)/\sqrt{2}$, but the phenomenology in the real singlet case is similar. 
 The vacuum stability of the real singlet model has previously been studied in \cite{Gonderinger:2009jp}. 
 
 Denoting the SM Higgs boson with $H_1,$ the most general Lagrangian invariant under the $Z_2$ transformations $H_1 \to H_1$, $S \to -S$  is
 given by 
\begin{equation}
\begin{split}
    V &= \mu_{1}^{2} \hc{H_{1}} H_{1} + \lambda_{1} (\hc{H_{1}} H_{1})^{2}
    + \mu_{S}^{2} \hc{S} S + \frac{\mu_{S}^{\prime 2}}{2} \left[ S^{2} + (\hc{S})^{2} \right]
    + \lambda_{S} (\hc{S} S)^{2} + \frac{ \lambda'_{S} }{2} \left[ S^{4} + (\hc{S})^{4} \right] \\
    &+ \frac{ \lambda''_{S} }{2} (\hc{S} S) \left[ S^{2} + (\hc{S})^{2} \right] + \lambda_{S1}( \hc{S} S) (\hc{H_{1}} H_{1}) + \frac{ \lambda'_{S1} }{2} (\hc{H_{1}} H_{1}) \left[ S^{2} + (\hc{S})^{2} \right].
\end{split}
\label{eq:Vsing}
\end{equation}
The vacuum stability conditions for the complex singlet model with a global $U(1)$ are given in \cite{Barger:2008jx}. However, those conditions are not 
applicable here because this model is far too simple compared to the general case \eqref{eq:Vsing}.  For the general model the full vacuum stability 
conditions are rather complicated and have been addressed previously in Ref.~\cite{Kadastik:2009cu}.  
However, the conditions of  \cite{Kadastik:2009cu} turn out to be too restrictive because they are derived by requiring the matrix of 
quartic couplings to be positive.  This is required only if the coefficients of biquadratic terms are negative and, in general, cut out some
allowed parameter space. 


The conditions arising from pure quartic terms of the potential \eqref{eq:Vsing} 
are
\begin{equation}
	\lambda_1 \geq 0, \quad \lambda_{S} + \lambda'_S \geq \abs{\lambda''_S}.
\label{eq:vacuum:stability:S:quartic}
\end{equation}
For simplicity we consider in addition only the case when the coefficents of the terms biquadratic in real fields (\emph{e.g.} the coefficient of $S_H^2 S_A^2$) are all non-negative, giving
\begin{equation}
	\lambda_S - 3 \lambda'_S \geq 0, \quad \lambda_{S1} 
	- \abs{\lambda'_{S1}} \geq 0.
\label{eq:vacuum:stability:S:all:biquadr:pos}
\end{equation}
Doing this, we exclude a part of the points that would be allowed by the full vacuum stability conditions. However, this is sufficient for our purposes because
our aim is  to show that  regions of the parameter space exist  that lower the SM Higgs boson mass vacuum stability bound.

\begin{figure}[t]
\centering
\includegraphics[width=0.5\textwidth]{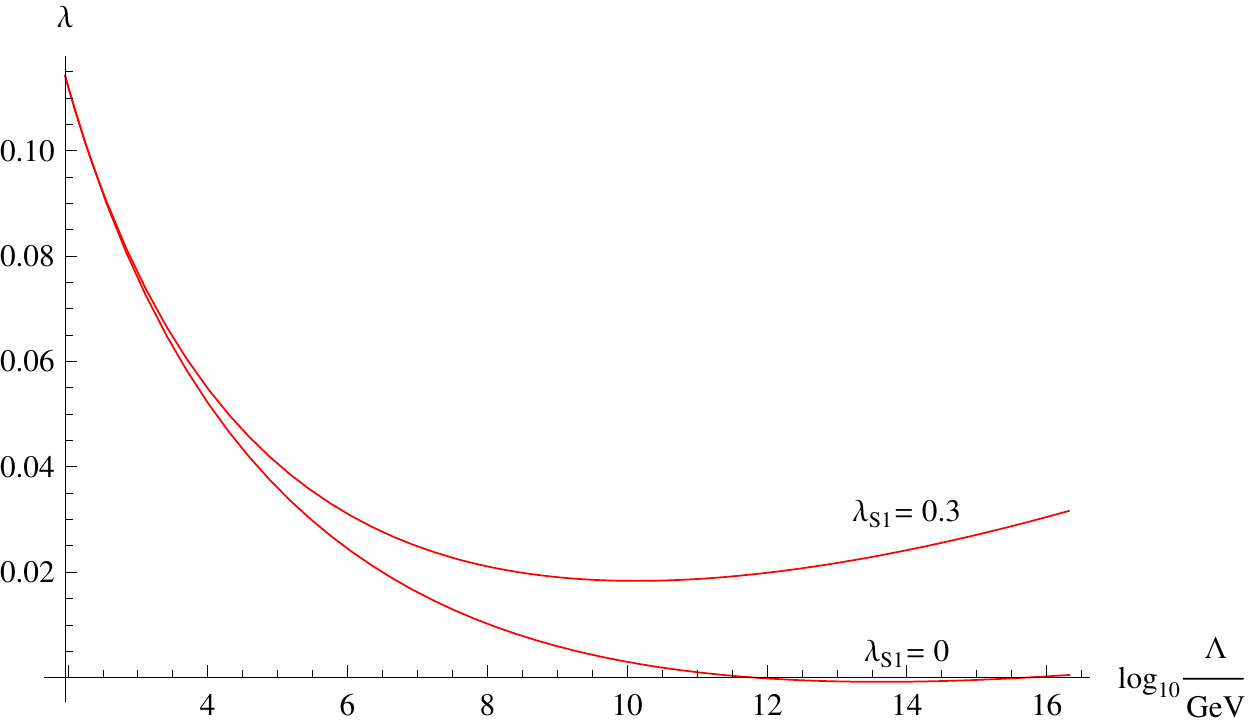}
\caption{Running of the Higgs self-coupling in the complex singlet model for two different values of $\lambda_{S1}$.}
\label{fig:lambdaS1:vs:MH}
\end{figure}

The one-loop RGEs can be obtained from those in \cite{Kadastik:2009cu} by setting all couplings of the inert doublet to zero.
The RGEs show that nonzero $\lambda_{S1}$ or $\lambda'_{S1}$ give a positive contribution to the $\beta$-function of $\lambda_1$, pushing the scale where $\lambda_1 \equiv \lambda = 0$ higher. For qualitative understanding of the model, we let $\lambda_S = \lambda'_S = \lambda''_S = \lambda'_{S1} = 0$. Fig.~\ref{fig:lambdaS1:vs:MH} shows one loop level running for the $125$~GeV Higgs quartic coupling for  $\lambda_{S1} = 0$ (the SM case) and for $\lambda_{S1} = 0.3$. In the latter case, the minimum bound on Higgs boson mass from the vacuum stability argument is lowered and the vacuum can be stable up to the GUT or Planck scale.

\subsection{ Inert doublet model}

In the inert doublet model \cite{LopezHonorez:2006gr,Barbieri:2006dq,Ma:2006km,Deshpande:1977rw} there is, besides the SM Higgs $H_1$, an additional scalar doublet $H_2$ that is odd under a new $Z_2$ symmetry and thus does not have Yukawa couplings. The neutral component of the inert doublet is a DM candidate. The most general Lagrangian invariant under the $Z_2$ transformations $H_1 \to H_1$, $H_2 \to -H_2$ is
\begin{equation}
\begin{split}
  V=&\mu_1^2|H_1|^2+\mu_2^2|H_2^2|+\lambda_1|H_1|^4+\lambda_2|H_2|^4+\lambda_3|H_1|^2|H_2|^2\\
    &+\lambda_4|H_1^\dagger H_2|^2+\frac{\lambda_5}{2}\left[(H_1^\dagger H_2)^2+\mathrm{h.c.}\right]\,.
\end{split}
\label{eq:Vinertdoublet}
\end{equation}

The requirement of vacuum stability imposes
\begin{equation}
\lambda_1,\lambda_2 > 0\,,\qquad
\lambda_3, \lambda_3+\lambda_4-|\lambda_5|>-2\sqrt{\lambda_1\lambda_2}\,.
\label{eq:vacstability:ID}
\end{equation}

We will not perform a detailed study of the inert doublet model alone here, because it is a limiting case of the singlet plus doublet model studied below.

\subsection{ Singlet plus doublet model}

This model has been previously studied in the context of $SO(10)$ GUT \cite{Kadastik:2009dj,Kadastik:2009cu,Kadastik:2009ca,Kadastik:2009gx,Huitu:2010uc}. Here, however, we present a general scan of parameters without imposing any GUT boundary conditions.

The Lagrangian with $Z_{2}$ even $H_{1}$ and odd $H_{2}$ and $S$ is
\begin{equation}
\begin{split}
    V &= \mu_{1}^{2} \hc{H_{1}} H_{1} + \lambda_{1} (\hc{H_{1}} H_{1})^{2}
    + \mu_{2}^{2} \hc{H_{2}} H_{2} + \lambda_{2} (\hc{H_{2}} H_{2})^{2}
    \\
    &+ \mu_{S}^{2} \hc{S} S + \frac{\mu_{S}^{\prime 2}}{2} \left[ S^{2} + (\hc{S})^{2} \right]
    + \lambda_{S} (\hc{S} S)^{2} + \frac{ \lambda'_{S} }{2} \left[ S^{4} + (\hc{S})^{4} \right]
    + \frac{ \lambda''_{S} }{2} (\hc{S} S) \left[ S^{2} + (\hc{S})^{2} \right] \\
    &+ \lambda_{S1}( \hc{S} S) (\hc{H_{1}} H_{1}) + \lambda_{S2} (\hc{S} S) (\hc{H_{2}} H_{2}) \\
    &+ \frac{ \lambda'_{S1} }{2} (\hc{H_{1}} H_{1}) \left[ S^{2} + (\hc{S})^{2} \right]
    + \frac{ \lambda'_{S2} }{2} (\hc{H_{2}} H_{2}) \left[ S^{2} + (\hc{S})^{2} \right] \\
    &+ \lambda_{3} (\hc{H_{1}} H_{1}) (\hc{H_{2}} H_{2})
    + \lambda_{4} (\hc{H_{1}} H_{2}) (\hc{H_{2}} H_{1})
    + \frac{\lambda_{5}}{2} \left[(\hc{H_{1}} H_{2})^{2} + (\hc{H_{2}} H_{1})^{2} \right] \\
    &+ \frac{\mu_{S H}}{2}  \left[\hc{S} \hc{H_{1}} H_{2} + \hc{H_{2}} H_{1} S \right]
    + \frac{\mu'_{S H}}{2}  \left[S \hc{H_{1}} H_{2} + \hc{H_{2}} H_{1} \hc{S} \right].
\end{split}
\label{eq:V:CP:inv}
\end{equation}

Just as for the complex singlet model, we consider here only the case of positive biquadratic terms for real fields (with the exception of the purely inert doublet conditions that are completely general). The simplified vacuum stability conditions for this model are given by \eqref{eq:vacuum:stability:S:quartic}, \eqref{eq:vacuum:stability:S:all:biquadr:pos} and \eqref{eq:vacstability:ID}
together with an additional constraint\footnote{Again, similarly to the singlet model the constraints in Ref.~\cite{Kadastik:2009cu} 
that were used in the previous version of the current paper are too restrictive.}
\begin{equation}
\lambda_{S2} - \abs{\lambda'_{S2}} \geq 0.
\label{eq:vacuum:stability:SIID:biquadr}
\end{equation}

\begin{figure}[t]
\includegraphics[width=0.49\textwidth]{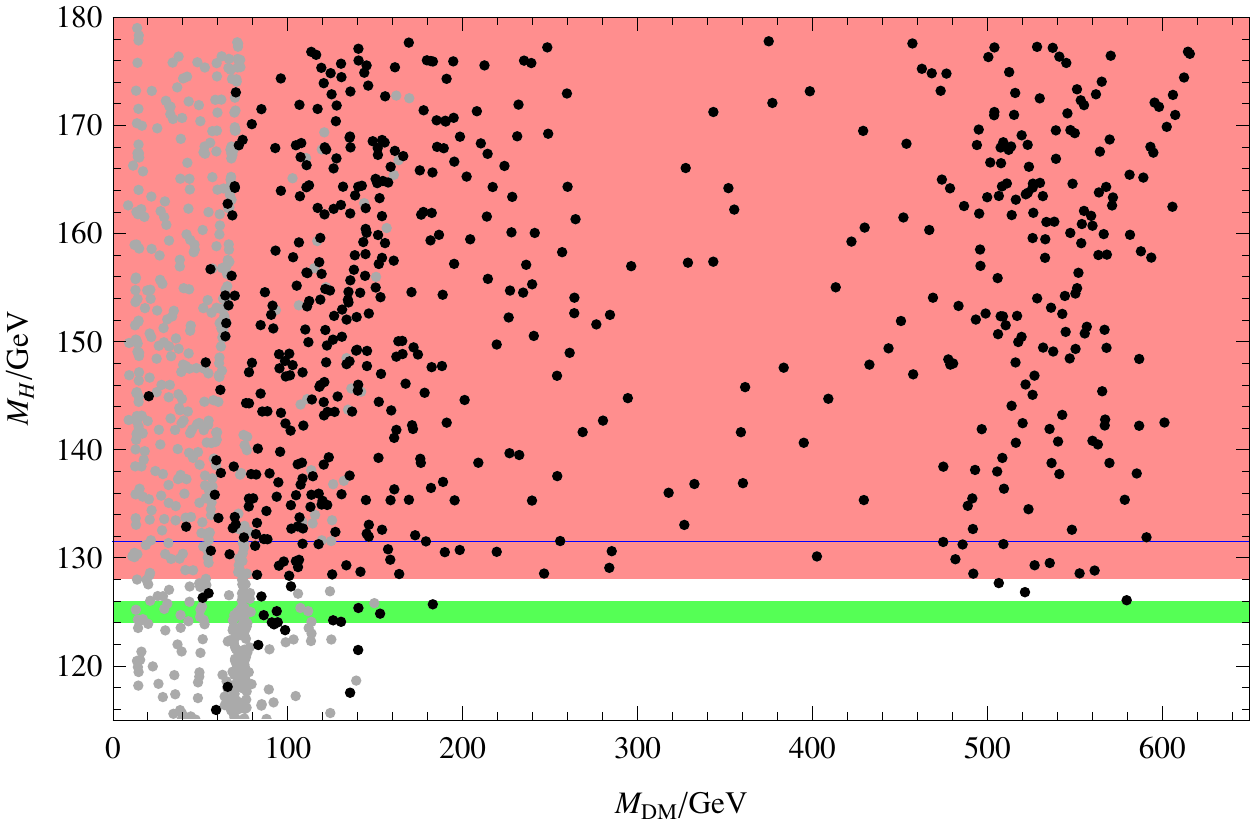}
\includegraphics[width=0.51\textwidth]{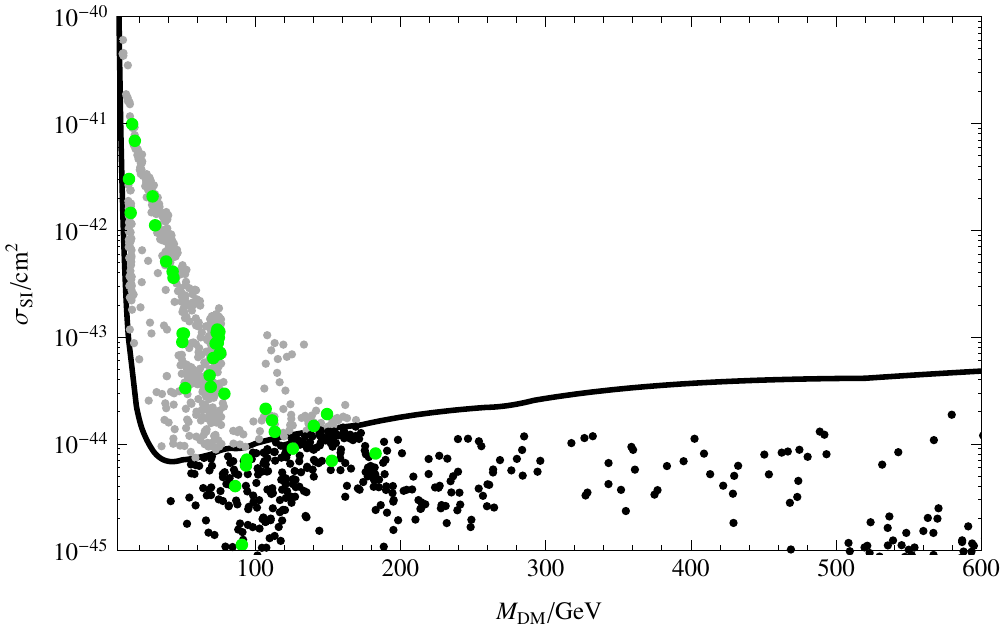}	
\caption{Left: Scatter plot of the Higgs boson mass predictions in the singlet plus doublet DM model at one loop level.
The blue line shows the SM one loop vacuum stability bound $M_H>131.5$~GeV for a  fixed 
$\Lambda_{\rm GUT}=2.1\times 10^{16}$~GeV. The light red area is excluded by the CMS, the green area shows $125\pm 1$~GeV. Gray points are excluded by the XENON100 bound, black points satisfy  the XENON100 bound.
Right: Dark matter spin-independent cross section $\sigma_{\rm SI}$ vs. DM mass. The black line is the XENON100 bound. Green points have $M_H$ in the $125\pm1$~GeV range.
}
\label{fig:singlet:doublet}
\end{figure}

The RGE-s for couplings and mass parameters are given in \cite{Kadastik:2009cu}.
We have performed a scan of the parameters for the values of couplings randomly generated in the ranges
\begin{equation}
\begin{aligned}
115~{\rm GeV} \leq M_H & \leq 180~{\rm GeV}, & 10~{\rm GeV} \leq \mu_S & \leq 10^3~{\rm GeV}, \\
10~{\rm GeV} \leq \mu_2 & \leq 10^3~{\rm GeV}, & 10~{\rm GeV^2} \leq \mu_S^{\prime 2} & \leq 100~{\rm GeV^2}, \\
10^{-2}~{\rm GeV} \leq \abs{\mu'_{SH}} & \leq 10^3~{\rm GeV}, & 0 \leq \lambda_2 & \leq 0.1, \\
0 \leq \lambda_S & \leq 0.1, & -0.1 \leq \lambda'_S & \leq 0.1, \\
-1 \leq \lambda_3  & \leq  1, & -1 \leq \lambda_4 & \leq  1, \\ 
0 \leq \lambda_{S1} & \leq 1, & 0 \leq \lambda_{S2} & \leq 1,
\end{aligned}
\end{equation}
with the rest of the parameters set to zero. In the case of every generated point we check that it satisfies  the requirements of vacuum stability and perturbativity in the whole range from $M_Z$ to $\Lambda_{\rm GUT}$, positivity of masses at $M_Z$ and lie within the $3 \sigma$ range of the WMAP cosmic abundance. The points that satisfy all the constraints are shown in Fig.~\ref{fig:singlet:doublet}.

In the left panel of Fig.~\ref{fig:singlet:doublet}, the region excluded by the CMS is shown in red; the $124-126$~GeV Higgs mass range is shown in green. Because the points were calculated using one-loop RGEs for the doublet plus singlet model, we show the GUT scale vacuum stability bound for the SM at one-loop level with the blue line (the two-loop bound is lower by about $3.5$~GeV). The points excluded by the XENON100 experiment~\cite{Aprile:2011hi} are shown in gray while the black points satisfy the present direct detection constraints. The shortage of points in the range from about $100$~GeV to about $500$~GeV is due to the DM being mostly singlet-like: in the low mass range it annihilates via the Higgs resonance, in the mass range above $500$~GeV the quartic scalar interactions can be large enough to allow for efficient annihilation via contact terms, but in between annihilation is not efficient, resulting in overaboundance of DM and exclusion by CMB bounds.

The right panel of Fig.~\ref{fig:singlet:doublet} shows the XENON100 direct detection constraints in detail. The points in the Higgs boson mass range $M_H = 125\pm1$~GeV are green. The low mass region below $50$~GeV is excluded. Between $100$~GeV and $200$~GeV there is a region that accommodates $M_H = 125\pm1$~GeV, having vacuum stability up to the GUT scale  with a low mass Higgs. Thus we conclude that the scalar DM models are perfectly consistent with the 125~GeV higgs mass and do not require the existence of new fundamental scale below the GUT or Planck scale. 

The scan is no exhaustive, but for 124-126 GeV Higgs mass range, the noticeable differences with the rest of the parameter space are in the soft coupling $\mu'_{SH}$ and couplings between dark sector and the SM Higgs that tend to be smaller than with a freely varying Higgs mass.

\section{CMSSM dark matter and LHC phenomenology for the 125~GeV Higgs boson}
\label{sec:cmssm}

The CMSSM is the most thoroughly studied SUSY model. Naturally, if the Higgs boson is discovered with
the mass $M_H=125\pm 1$~GeV, one would like to know what is the implication of this discovery 
for the phenomenology of this model. Here we show that if all the phenomenological constraints are taken into account,
the CMSSM parameter space shrinks into well defined small regions according to the dominant DM freeze-out
process. We study whether the CMSSM can be tested at the LHC and in DM direct detection experiments such as XENON100
and conclude that, despite of heavy Higgs boson, discovery of CMSSM gluinos and/or stops is not excluded at the LHC.
In addition, if the sparticle spectrum is too heavy for the LHC discovery, DM direct detection experiments may still discover the CMSSM DM. 

It is well known that such a heavy Higgs boson imposes challenges on SUSY models in which the Higgs boson mass is 
predicted to be
\bea
M_H^2=M_Z^2\cos^2 2\beta + \delta_t^2,
\eea
where $\delta_t$ is the stop dominated loop contribution. For $M_H\approx 125$~GeV the loop contribution must be
as large as the tree level one which requires very heavy stops unless there is an extremely  large trilinear scalar coupling
that makes the lightest stop light due to  large mixing.
A heavy SUSY scale, in turn, makes the lightness of electroweak symmetry breaking scale unnatural. In addition,
a heavy sparticle spectrum imposes fine tunings on the processes that contribute to the DM freeze-out in SUSY models.
Taking those facts into account, the  phenomenological constraints that are commonly addressed in the context
of SUSY models, summarized in Table~\ref{tab}, the constraints from SUSY searches at the LHC and
the constraints from DM direct detection, the CMSSM parameter space is known to be rather 
fine tuned~\cite{Farina:2011bh,Buchmueller:2011ki,Buchmueller:2011sw,Bertone:2011nj,Fowlie:2011mb}. 

At the GUT scale the parameter space of the CMSSM is described by five parameters,
\bea
m_0,\; M_{1/2},\; A_0, \; \tan\beta,\; {\rm sign} (\mu),
\eea
the common scalar mass, the common gaugino mass, the common trilinear coupling, ratio of two Higgs vevs and the 
sign of the higgsino mass parameter. To scan over the CMSSM parameter space we randomly generate the parameters 
in the following ranges:
\begin{equation}
\begin{aligned}
300 < & m_0, &  & M_{1/2} < 10^4~{\rm GeV}, \\ 
|A_0| < & 5 m_0, & 3< & \tan\beta < 60, \\
{\rm sign}(\mu)=&\pm.
\end{aligned}
\end{equation}
We use the MicrOMEGAs package \cite{Belanger:2010gh,Belanger:2006is} to compute the electroweak scale sparticle mass spectrum, the Higgs boson masses, the DM relic abundance $\Omega_{\rm DM}$,
the spin-independent DM-nucleon direct detection cross section $\sigma_{\rm SI}$ and the other observables in Table~\ref{tab}.
In addition, we require $M_H=125\pm 1$~GeV. We do not attempt to find the best fit regions of the parameter space because
there is no Higgs mass measurement yet. In addition, there is a few GeV theoretical uncertainty in the computation of SUSY Higgs masses
in the available codes. 
Therefore, to select the phenomenologically acceptable parameter space  we impose $3\sigma$ hard cuts for the observables in  Table~\ref{tab}.%
\footnote{The new constraints on $B_s \to \mu^+\mu^-$ from the LHCb and CMS \cite{Aaij:2012ac,CMSbstomumu} have an impact on points with low stop mass at
high $\tan \beta$. Qualitatively, however, the regions and channels remain the
same.}
Our approach should be regarded as an example study of the CMSSM parameter space
for heavy Higgs boson; qualitatively similar results should hold if the real Higgs boson mass deviates from 125~GeV by a few GeV.

\begin{table}[t]
\begin{center}
\begin{tabular}{c|cc}
Quantity & Experiment & Standard Model\\ \hline
$\alpha_3(M_Z)$~\cite{Bethke:2009jm} &  $0.1184 \pm 0.0007$ & parameter\\
$m_t$~\cite{Lancaster:2011wr} & $173.2\pm 0.9$&parameter\\ 
$m_b$~\cite{PDG} & $4.19\pm 0.12$&parameter\\ 
$\Omega_{\rm DM} h^2$~\cite{Larson:2010gs} & $0.112 \pm 0.0056$ & 0\\
$\delta a_\mu$~\cite{Davier:2010nc} & $(2.8\pm0.8)\times10^{-9}$ & 0\\
BR$(B_d\to X_s\gamma)$~\cite{Misiak:2006zs} & $(3.50\pm0.17)\times10^{-4}$ & $(3.15\pm0.23)\,10^{-4}$ \\
BR$(B_s\to \mu^+\mu^-)$~\cite{Bsmumu} & $<1.1\times10^{-8}$ at 95\%C.L.& $(0.33\pm 0.03)\,10^{-8}$\\
BR$(B_u\to \tau\bar\nu)$/SM~\cite{Buchmueller:2009fn} & $1.25\pm0.40$ & 1\\
\end{tabular}
\end{center}
\caption{\label{tab}\em Used constraints for the CMSSM analyses.}
\end{table}

\begin{figure}[t]
\includegraphics[width=\textwidth]{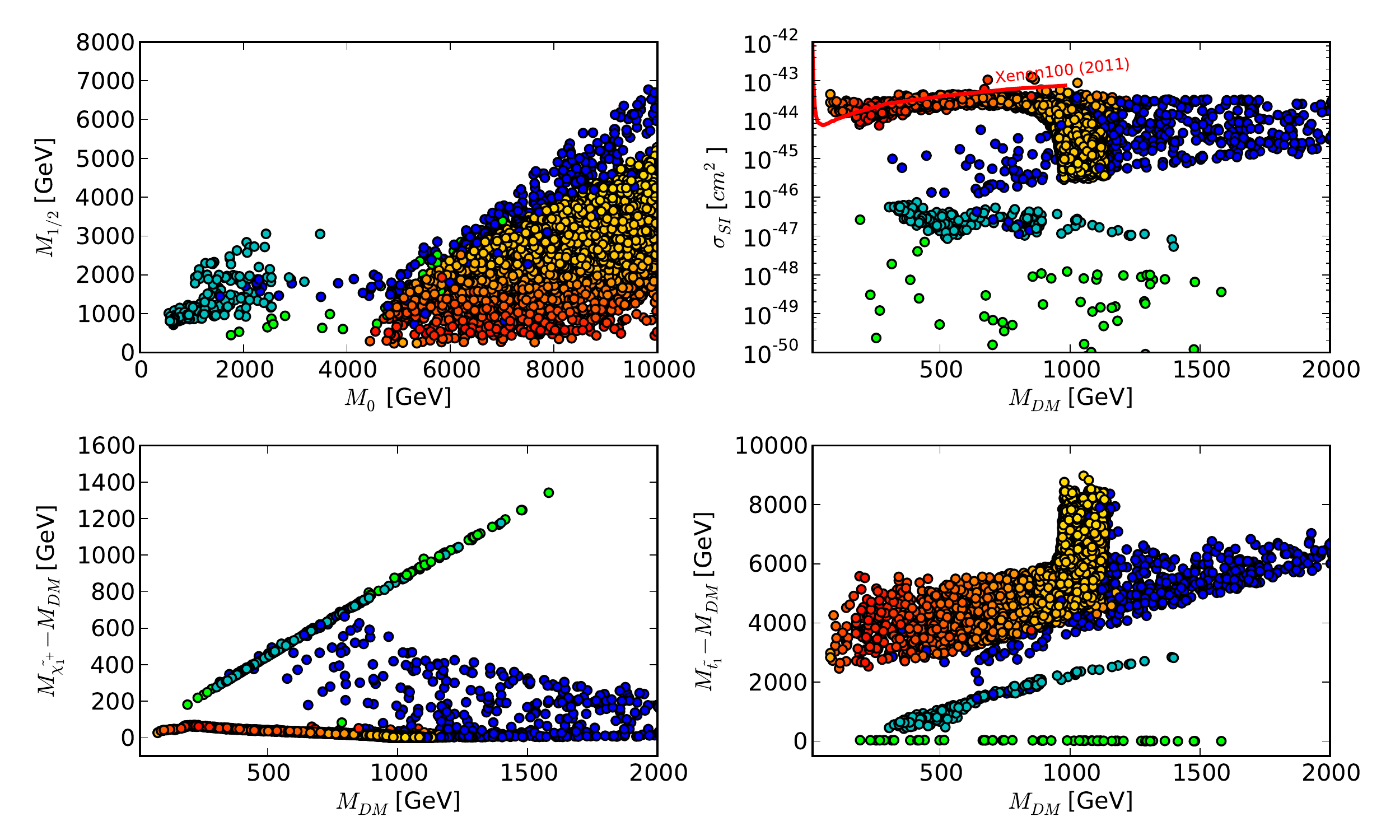}
\caption{
Scatter plots over the CMSSM parameter space keeping $M_H=125\pm 1$~GeV. Colours represent different 
dominant DM freeze-out processes.
Light blue: slepton co-annihilation; green: stop co-annihilation; red to orange: well-tempered neutralino,
yellow: higgsino;  dark blue: heavy Higgs resonances. No $(g-2)_\mu$ constraint is imposed.}
\label{fig2}
\end{figure}

Our results are presented in Figs.~\ref{fig2}-\ref{fig:cmssm:g-2}.
 Because there is a tension between the observables that push the SUSY scale
to high values and the measurement of  $(g-2)_\mu$~\cite{Farina:2011bh},  we disregard the $(g-2)_\mu$ constraint for the moment.
The reason is that the CMSSM parameter fit is largely dominated by two observables, the DM relic abundance and the $(g-2)_\mu$,
the latter constraining mostly the scale. We would first like to study the parameter space that induces correct $M_H$ and $\Omega_{\rm DM}$.
Therefore we discuss the implications of the $(g-2)_\mu$ constraint later.

In Fig.~\ref{fig2} we present our results in scatter plots without the $(g-2)_\mu$ constraint.
In the upper left panel the results are presented in $(m_0, M_{1/2})$ plane,
in the upper right panel  in $(M_{\rm DM}, \sigma_{\rm SI})$ plane,
in the lower left panel  in $(M_{\rm DM}, M_{\tilde \chi_1^+}-M_{\rm DM})$ plane, and 
in the lower right panel  in $(M_{\rm DM}, M_{\tilde t_1}-M_{\rm DM})$ plane.
The first 100 days XENON100 constraint~\cite{Aprile:2011hi} is also shown.

We identify five distinctive parameter regions according the dominant DM annihilation processes.
\begin{itemize}
\item The light blue points with small  $m_0$ and $ M_{1/2}$
represent the slepton co-annihilation region. They  are featured by very large values of $\tan\beta.$
Those points represent the best fit value of the CMSSM~\cite{Farina:2011bh}
and have low enough sparticle masses that allow potential SUSY discovery at the LHC. However, their spin-independent direct detection cross section
is predicted to be below $10^{-46}$~cm$^2$ and remains unobservable at the XENON100. The present XENON100 experimental bound is 
plotted in the upper right panel with solid red line. This is the only parameter region that survives at $3\sigma$ level after the  $(g-2)_\mu$ constraint is
 imposed.

\item The green dots  represent the stop co-annihilation region. Consequently those points have the lowest possible stop mass and, due to the 
mass degeneracy with DM, stops can be long lived and seen as stable very slow particles ($R$-hadrons) at the LHC. 
The feature of those points is an enormous trilinear coupling and very large stop mixing. In addition, the gluino mass can be 
reachable at the LHC. For stop co-annihilation region the  spin-independent DM direct detection cross section
is, unfortunately,  unobservable.
\begin{figure}[t]
\includegraphics[width=\textwidth]{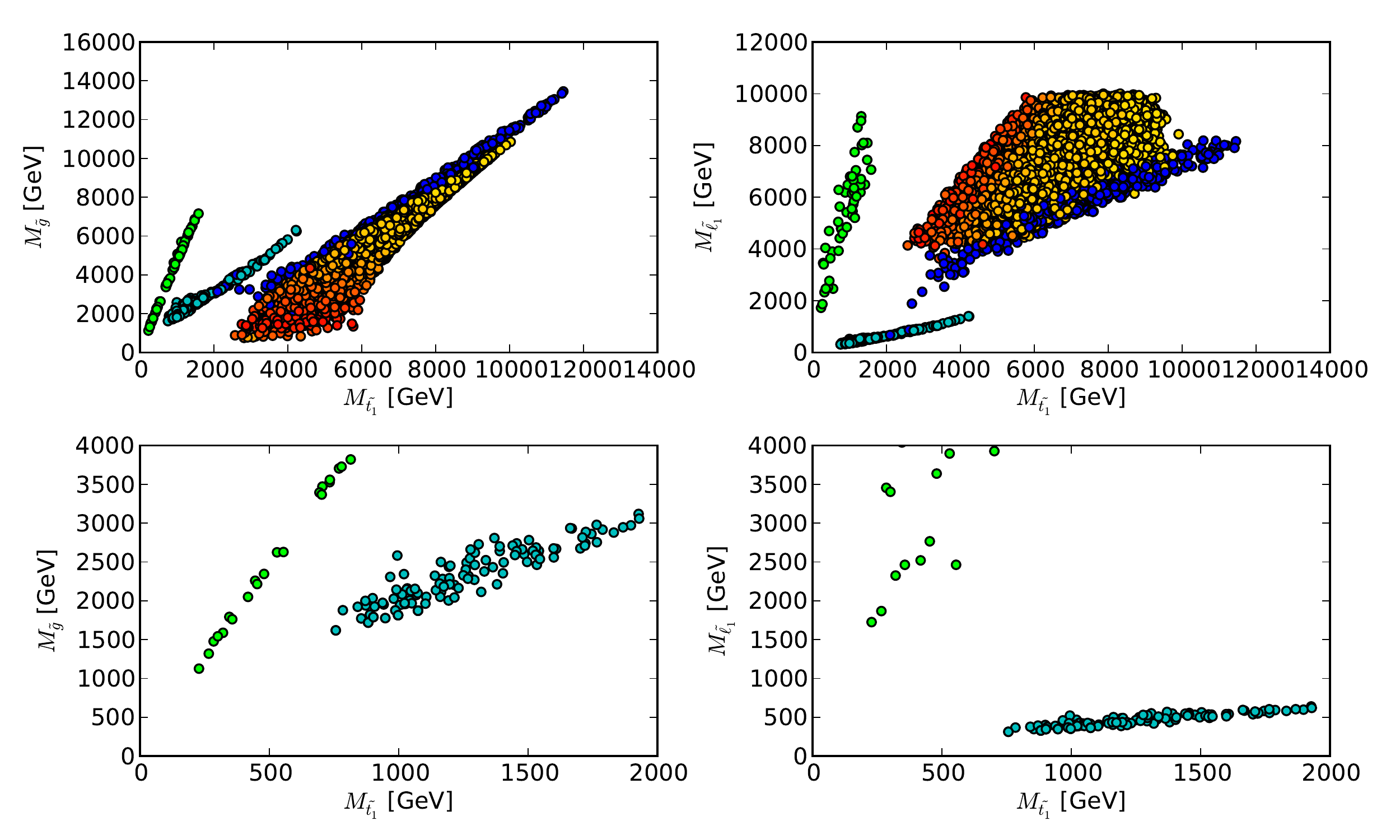}
\caption{
The same as in Fig.~\ref{fig2} but for physical gluino and the lightest stop and slepton  masses. The lower panels present low mass
zoom to the upper panels.}
\label{fig3}
\end{figure}

\item The dots represented by continuous colour code from red to orange 
represent the so called well-tempered neutralino~\cite{ArkaniHamed:2006mb}, \ie, neutralinos with large bino-higgsino mixing.
The colour varies according to the higgsino component from red (predominantly bino) to yellow (pure higgsino).
Therefore those points can simultaneously have small DM mass  and large DM-nucleon scattering cross section 
 that can be well tested at the XENON100. However, apart from the DM,
all other sparticle masses are predicted to be too heavy for direct production at the LHC.

\item The yellow dots around $M_{\rm DM}\sim 1$~TeV represent the pure higgsino DM that is almost degenerate in mass with chargino.
 The sparticle mass spectrum is predicted to be even heavier than 
in the previous case because the DM scale is fixed to be high.
These points represent the most general and most abundant bulk of the  $M_H=125$~GeV Higgs scenario --
apart from the light DM and heavy Higgs boson there are no other observable consequences because stops can completely decouple.
In our case the 10~TeV bound on stops is imposed only because we did not generate larger values of $m_0.$

\item The dark blue points represent heavy Higgs resonances. Those points are featured by very large values of $\tan\beta$ and  give the heaviest mass spectrum. In essence those points are just smeared out higgsino points due to additional Higgs-mediated processes.
\end{itemize}

In order to study the testability of those parameter regions at the LHC we plot in Fig.~\ref{fig3} 
the physical gluino mass against the lightest stop mass and the lightest slepton mass against the lightest stop mass.
Clearly, the only two regions of interest for the LHC are the slepton and stop co-annihilation regions. Therefore
we plot in lower panels the low mass scale zoom of the upper panels. According to Ref.~\cite{Baer:2011aa}
both regions have a chance to be discovered already in the 7~TeV LHC. Interestingly, due to the stop mass degeneracy with DM
the stops can be long-lived. In this case one must search for $R$-hadrons at the LHC experiments.

To study the $\tan\beta$ and heavy Higgs mass dependence of the generated parameter space we plot in 
 Fig.~\ref{fig4} scatter plots in $(m_{\tilde t_1},\tan \beta)$ and  $(M_{A},\tan \beta)$ plains. The slepton
 co-annihilation points have a preferably large $\tan\beta$ that implies large contributions to the observables 
 like $B_s\to \mu \mu$ and the $(g-2)_\mu.$ Those allow for indirect testing of this parameter region. Unfortunately the 
 heavy Higgses are predicted to be too heavy to detect at the LHC.

\begin{figure}[t]
\includegraphics[width=1.\textwidth]{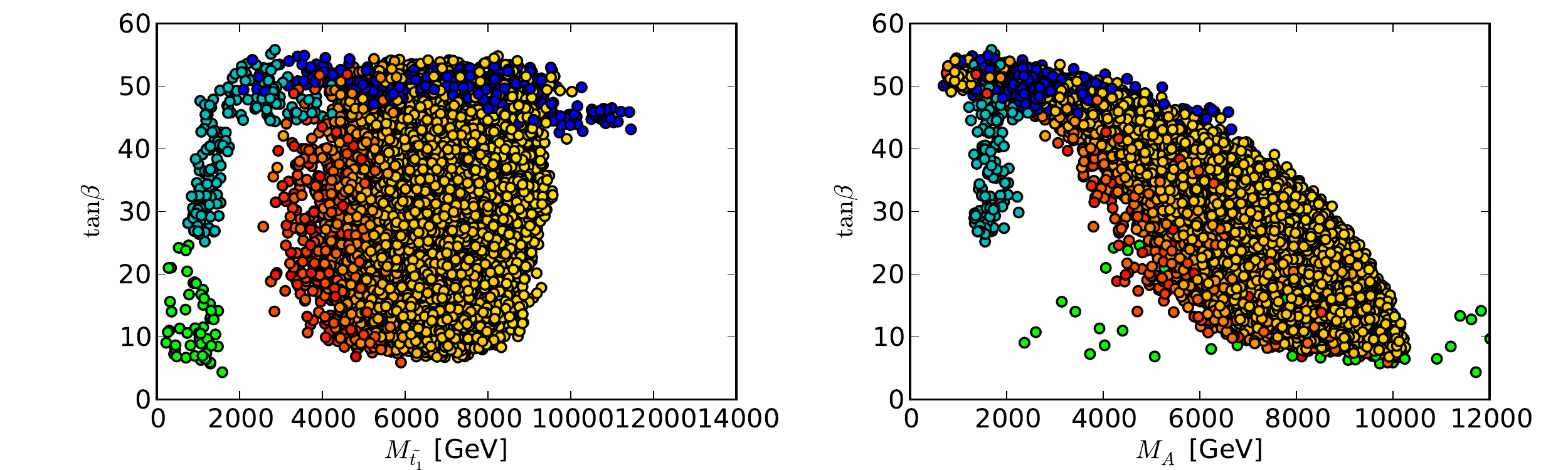}
\caption{
The same as in Fig.~\ref{fig2} but in $(M_{\tilde t_1},\tan\beta)$ and $(M_{A},\tan\beta)$ planes.}
\label{fig4}
\end{figure}

We remind that so far we have disregarded the $(g-2)_\mu$ constraint. If we impose a hard $3\sigma$ cut 
on the generated parameter space, only the slepton co-annihilation region survives. The result is plotted in Fig.~\ref{fig:cmssm:g-2}
where we repeat the content of  Fig.~\ref{fig2} but with the additional $(g-2)_\mu$ constraint. As expected, the observed deviation
in the $(g-2)_\mu$ from the SM prediction is hard to explain in SUSY models with heavy spectrum. 
Therefore the two measurements,  $(g-2)_\mu$ and $M_H=125$~GeV,  are in conflict in the CMSSM \cite{Baer:2011ab}. 
The conflict is mildest in the slepton co-annihilation case
because of large $\tan\beta$ and the lightest sparticle spectrum. 
 Therefore, for the $M_H=125$~GeV Higgs boson, we predict definite sparticle masses and correlations between them, 
shown in  Fig.~\ref{fig:cmssm:g-2}, for the LHC. If the CMSSM is realized in Nature and if it contributes significantly to the $(g-2)_\mu$,
 the sparticle spectrum is essentially fixed and potentially observable at the LHC.

\begin{figure}[t]
\includegraphics[width=1.\textwidth]{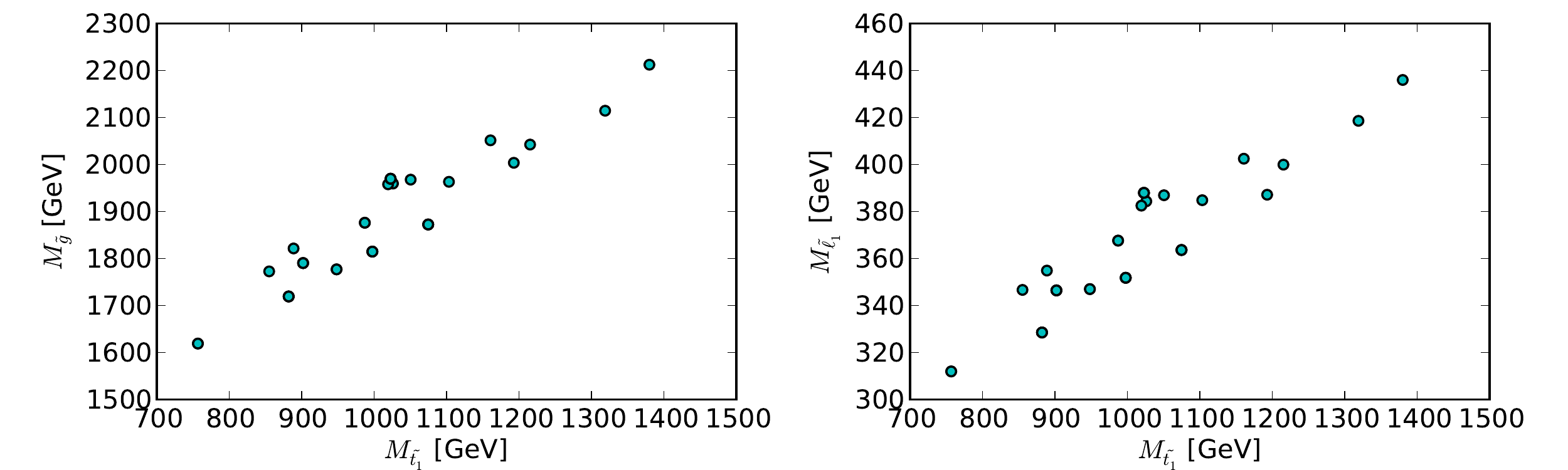}
\caption{The same as  in Fig.~\ref{fig3} in the case of imposing $3\sigma$ constraint on the $(g-2)_\mu$ prediction.}
\label{fig:cmssm:g-2}
\end{figure}

\section{Conclusions}
\label{sec:concl}

The recent LHC searches for the SM-like Higgs boson motivate studies of the fundamental scalars  in particle physics models and in cosmology.
In this paper we analyzed the implications of  the $M_H\approx 125$~GeV Higgs boson for the vacuum stability in scalar DM models and
for the phenomenology of CMSSM. This value of the Higgs boson mass is interesting in both cases because it does not fit to the
standard expectation neither in the SM nor in minimal supersymmetric  models with SUSY breaking scale below 1~TeV. 

We have shown that in the case of non-SUSY scalar DM models  the vacuum can be stable up to the GUT scale even for Higgs boson masses much
 below the corresponding SM bound. Therefore, unlike the SM,  
 the scalar DM models can be valid up to the GUT or Planck scales even for the Higgs boson mass as low as $M_H\approx 125$~GeV.

In minimal SUSY models, to the contrary, the   $M_H\approx 125$~GeV Higgs boson is heavier than expected in scenarios that address 
naturalness of the electroweak scale. In order to generate such a large Higgs boson mass at loop level, the SUSY breaking scale must be rather high
and could be unobservable at the LHC. This problem can be overcome with extremely large stop $A$-term so that the lightest stop 
is light due to large mixing. At the same time the DM neutralino can also be light, either because of dominant slepton co-annihilation processes
or because of large bino-higgsino mixing. In the latter case the DM-nucleon scattering cross section can be observable in 
direct detection experiments like the XENON100.

To quantify those results we studied the CMSSM by scanning over its parameter space  allowing the sparticle 
mass parameters to be very large. We first considered the case without attempting to explain the $(g-2)_\mu$ in the context of CMSSM.
 We confirmed that for very large $A$-terms there exists a stop co-annihilation region 
where all DM,  stop and gluino are preferably light. Due to the mass degeneracy between stop and DM the stops can also be long lived
resulting in non-trivial LHC phenomenology. The second parameter region that is potentially reachable at the LHC is the slepton
co-annihilation region. The most important result of this work is to make sharp predictions of gluino, stop and slepton masses,
shown in Fig.~\ref{fig3},  for the CMSSM parameter regions that remain testable at the LHC.

For other channels of generating the correct DM  relic abundance 
the $M_H\approx 125$~GeV Higgs boson implies very heavy sparticle masses. The exception is, of course,
the DM that can be light due to bino-higgsino mixing even if other sparticles are as heavy as 10~TeV. In this case the CMSSM 
cannot be tested at the LHC but  the DM spin-independent scattering cross section off nuclei may be large due to the
large higgsino component. The latter scenario may be discoverable already in the running XENON100 experiment. 

If, in addition, one attempts to explain  also the $(g-2)_\mu$ in this framework, there is immediate tension between the high SUSY scale and the 
large value of the needed $(g-2)_\mu$ contribution. We found that after  imposing the $(g-2)_\mu$ constraint on the CMSSM, only the slepton
co-annihilation region survived at $3\sigma$ level, see Fig.~\ref{fig:cmssm:g-2}. 
This implies that the CMSSM has a definite prediction for the sparticle masses and spectrum
to be tested at the LHC experiments.

\vskip 0.5in
\vbox{
\noindent{ {\bf Acknowledgements} } \\
\noindent  
We thank A. Strumia for several discussions.
This work was supported by the ESF grants  8090, 8499, 8943, MTT59, MTT60, MJD140, JD164,  by the recurrent financing SF0690030s09 project
and by  the European Union through the European Regional Development Fund.
}


\bibliographystyle{JHEP} 
\bibliography{higgs_cmssm}

\end{document}